\documentclass[draft,twoside,leqno,twocolumn]{article}

\usepackage{siamproceedings}
\usepackage[T1]{fontenc}
\usepackage[utf8]{inputenc}
\usepackage{tikz}
\usepackage{pgfplots}
\usetikzlibrary{arrows.meta,calc,positioning,shapes.geometric,decorations.pathreplacing,fit,matrix,trees}
\pgfplotsset{compat=1.18}
\usepackage{amsmath,amssymb,amsfonts}
\usepackage{enumitem}
\usepackage{mathtools}
\usepackage{graphicx}
\usepackage{amsfonts}
\usepackage{booktabs}
\usepackage{url}
\usepackage{etoolbox}
\usepackage{tabularx}
\usepackage{adjustbox}
\usepackage{hyperref}
\usepackage{authblk}
\usepackage[yyyymmdd]{datetime}

\usepackage{subcaption}
\usepackage{multirow}
\usepackage{placeins}
\usepackage{cleveref}
\usepackage{supertabular}
\usepackage{scalerel}

\newcommand{\secref}[1]{Section~\ref{sec:#1}}

\newcommand{\tabref}[1]{Table~\ref{tab:#1}}

\newcommand{\figref}[1]{Figure~\ref{fig:#1}}

\def\cellsize{0.12}

\newcommand{\bc}[2]{
    \pgfmathsetmacro{\x}{((#2)-1)*\cellsize}
    \pgfmathsetmacro{\y}{-((#1)-1)*\cellsize}

      \fill[
        gray!65
      ]
      ($(M)+(\x,\y)$)
      rectangle
      ++(\cellsize,-\cellsize);
}

\newcommand{\wc}[2]{
    \pgfmathsetmacro{\x}{((#2)-1)*\cellsize}
    \pgfmathsetmacro{\y}{-((#1)-1)*\cellsize}

      \draw[
        gray!35,
        line width=0.1pt
      ]
      ($(M)+(\x,\y)$)
      rectangle
      ++(\cellsize,-\cellsize);
}

\newcommand{\hgridline}[2]{
    \pgfmathsetmacro{\x}{((#2)-1)*\cellsize}
    \pgfmathsetmacro{\y}{-((#1)-1)*\cellsize}

      \draw[
        gray!85,
        line width=.75pt
      ]
      ($(M)+(\x,\y)$)
      --
      ($(M)+(\x,-50*\cellsize)$);
}

\newcommand{\vgridline}[2]{
    \pgfmathsetmacro{\x}{((#2)-1)*\cellsize}
    \pgfmathsetmacro{\y}{-((#1)-1)*\cellsize}

      \draw[
        gray!85,
        line width=.75pt
      ]
      ($(M)+(\x,\y)$)
      --
      ($(M)+(50*\cellsize,\y)$);
}

\newcommand{\level}[1]{
  \hgridline{#1}{#1}
  \vgridline{#1}{#1}
}

\newcommand{\pgrid}{
  \foreach \i in {15,21,27,33,39,45} {
    \hgridline{9}{\i}
    \vgridline{\i}{9}
  }
}

\newcommand{\pidentity}[2]{
  \foreach \i in {1,...,6} {
    \bc{9 + 6*(#1) + \i - 1}{9 + 6*(#2) + \i - 1}
  }
}

\newcommand{\pidentities}{
  \foreach \j in {1,...,6} {
    \pidentity{0}{\j}
    \pidentity{\j}{0}
  }
}

\newtoggle{anonymous} 

\togglefalse{anonymous}

\title{Less is Moore}

\date{}

\iftoggle{anonymous} 
{ 
\author{}
}
{
\author[1]{Paul Stankovski Wagner}
\affil[1]{Lund University, Lund, Sweden, \texttt{{\small paul.stankovski\_wagner@eit.lth.se}}}
}

\newcommand{\popcount}{\operatorname{popcount}}

\begin{document}
\maketitle

\begin{abstract}
A Moore graph simultaneously asks for two things that normally pull in opposite directions: every vertex should have many neighbors, while every pair of vertices should remain close.  For diameter two and degree $k$, the Moore bound is $k^2+1$, and equality forces a remarkably rigid graph: it is $k$-regular, has girth five, or alternatively, is strongly regular with parameters $(k^2+1,k,0,1)$. Hoffman and Singleton showed that such a graph can exist only for $k=2,3,7$, and possibly $57$.  The first three cases are the pentagon, Petersen graph, and Hoffman--Singleton graph. The degree-$57$ case on 3,250 vertices remains open.

This paper is not an attempt to settle that open problem.  Instead, we ask a more algorithmic question: How efficiently can the defining Moore structure itself be searched exactly?  We begin from a human-developed exhaustive search based on the radius-two Moore tree.  The leaves fall into $k$ groups of size $k-1$, and the unknown edges between every pair of such groups form a perfect matching -- a permutation.  A partial construction is abandoned immediately when it creates a triangle or quadrangle. This simple observation already eliminates almost all arbitrary graphs from consideration.

We then describe a sequence of exact algorithm-engineering improvements developed through a human-guided AI process and accepted only after mathematical checking and controlled regression tests.  The main steps are reversible short-cycle forbid counters, bit-parallel domains, minimum-remaining-value branching, singleton propagation, matching-based \textsc{AllDifferent} filtering, balanced replayable parallel frontiers, residual-symmetry orbit pruning, and canonical deduplication of frontier states. On the principal $k=9$ benchmark, stronger symmetry reduces worker search from 7.29 million to 2.92 million nodes, cutting wall time from 47.6 to 17.9 seconds on a six-core laptop setup. Further parallelization speedup is reported separately; on a 16-core/32-thread machine, five repeated complete runs with 1,024 requested frontier jobs have a median wall time of 5.28 seconds. Degrees $10$ and $57$ are used as stress tests rather than completed searches.  A five-minute degree-$57$ run uses only 161 MiB peak resident memory, showing that state size is also modest.
\end{abstract}

\section{Introduction}

The degree/diameter problem asks a deceptively simple extremal question: How many vertices $n$ can a graph have if its maximum degree $k$ is fixed and every pair of vertices must be connected by a short path? Choose an arbitrary vertex $v_0$ as the root. With diameter two, there can be at most $k$ vertices at distance one from $v_0$, and each of these vertices $v_1,\ldots,v_k$ can introduce at most $k-1$ new vertices at distance two. Consequently
\begin{equation}
  n \le 1+k+k(k-1)=k^2+1 .
  \label{eq:moorebound}
\end{equation}
A (connected, undirected) graph meeting this elementary counting bound is a \emph{Moore graph} of diameter two. The bound is easy to derive; meeting it is exceptionally difficult. A broad account of the degree/diameter problem is given by Miller and \v{S}ir\'a\v{n}~\cite{miller2013moore}, while Dalf\'o~\cite{dalfo2019survey} surveys what is known specifically about the unresolved diameter-two case.

Equality in~\eqref{eq:moorebound} means that the radius-two breadth-first search around \emph{every} vertex contains no collisions.  Therefore a diameter-two Moore graph contains neither triangles nor quadrangles.  Conversely, a $k$-regular graph on $n=k^2+1$ vertices with no triangle or quadrangle has exhausted the entire Moore tree from every root and must have diameter two.

These Moore graphs can alternatively be characterized as being strongly regular with parameters
\begin{equation}
        (n,k,\lambda,\mu)=(k^2+1,k,0,1),
        \label{eq:srgparams}
\end{equation}
indicating that adjacent vertices have no common neighbor ($\lambda{}=0$), and non-adjacent vertices have exactly one ($\mu{}=1$). The $\lambda$ and $\mu$ conditions may, equivalently, be replaced with the condition that girth $g=5$.

These equivalent characterizations are be useful in the sequel, the one sporting non-existence of triangles and quadrangles in particular.

\begin{figure*}[ht]
\centering
\begin{tikzpicture}[scale=0.92, every node/.style={font=\scriptsize}]
\begin{scope}[xshift=0cm]
  \node[font=\bfseries] at (0,2.15) {$k=2$: $C_5$};
  \foreach \i in {0,...,4}{
    \coordinate (c\i) at ({90-72*\i}:1.35);
    \fill (c\i) circle (2pt);
  }
  \foreach \i/\j in {0/1,1/2,2/3,3/4,4/0}{\draw (c\i)--(c\j);}
  \node[align=center] at (0,-1.9) {$n=5$\\$(5,2,0,1)$};
\end{scope}
\begin{scope}[xshift=5.0cm]
  \node[font=\bfseries] at (0,2.15) {$k=3$: Petersen};
  \foreach \i in {0,...,4}{
    \coordinate (o\i) at ({90-72*\i}:1.5);
    \coordinate (i\i) at ({90-72*\i}:0.72);
    \fill (o\i) circle (2pt);
    \fill (i\i) circle (2pt);
    \draw (o\i)--(i\i);
  }
  \foreach \i/\j in {0/1,1/2,2/3,3/4,4/0}{\draw (o\i)--(o\j);}
  \foreach \i/\j in {0/2,2/4,4/1,1/3,3/0}{\draw (i\i)--(i\j);}
  \node[align=center] at (0,-1.9) {$n=10$\\$(10,3,0,1)$};
\end{scope}
\begin{scope}[xshift=10.6cm]
  \node[font=\bfseries] at (0,2.15) {$k=7$: Hoffman--Singleton};
  \node[draw,rounded corners,minimum width=3.7cm,minimum height=1.0cm,align=center] (P) at (0,0.8)
       {$P_0,\ldots,P_4$\\five pentagons};
  \node[draw,rounded corners,minimum width=3.7cm,minimum height=1.0cm,align=center] (Q) at (0,-0.75)
       {$Q_0,\ldots,Q_4$\\five pentagrams};
  \draw[-{Latex[length=2mm]},line width=.55pt] (P) -- node[right,align=left,font=\tiny] {$25$ cross matchings\\$j\mapsto hi+j\pmod 5$} (Q);
  \node[align=center] at (0,-2.05) {$n=50$\\$(50,7,0,1)$};
\end{scope}
\end{tikzpicture}
\caption{The three known diameter-two Moore graphs.  The Hoffman--Singleton panel shows its standard five-pentagon/five-pentagram construction rather than drawing all 50 vertices and 175 edges.  The compression is intentional: the important feature for the search developed here is not a particular drawing/embedding, but the increasing rigidity forced by $(n,k,\lambda,\mu)=(k^2+1,k,0,1)$.}
\label{fig:known-moore}
\end{figure*}
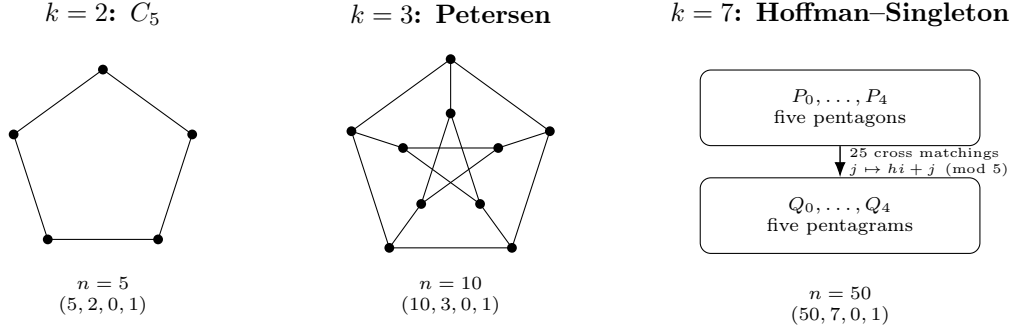

Hoffman and Singleton proved in~\cite{hoffman1960moore} that a diameter-two Moore graph can exist only for degrees
\begin{equation}
                   k\in\{2,3,7,57\},
\end{equation}
where the first three possibilities are realized uniquely and the last remains undecided. \figref{known-moore} shows the three known examples for $k=2,3,7$, with the last one in compressed form.  For $k=2$ we obtain the pentagon $C_5$.  For $k=3$ we obtain the Petersen graph.  For $k=7$ we obtain the 50-vertex Hoffman--Singleton graph.  A degree-$57$ Moore graph, if it exists, would contain 3,250 vertices and is commonly called the \emph{missing Moore graph} or the \emph{monster graph}.

\subsection{Motivation}
Why write software for a problem that has resisted mathematics for more than sixty years?  There are three reasons.  First, computational search is itself a way of learning which parts of the structure are difficult.  A search that repeatedly fails at the same local configurations can suggest useful lemmas, symmetries, or forbidden substructures.  Second, better exact software turns small and medium degrees into controlled laboratories.  We can measure what pruning rules really buy, which symmetries remain after normalization, and where the computation is spent.  Third, the missing graph is a demanding benchmark for exact combinatorial search.  A method useful here may transfer to other highly constrained designs, cages, incidence structures, permutation systems, and constraint-satisfaction problems.

The aim of this paper is therefore deliberately modest in its mathematical claim and ambitious in its algorithmic one.  We study how far a transparent exhaustive Moore-structure search can be improved while preserving exactness.  The development starts from a human-designed search whose central rule is almost embarrassingly simple: construct only edges that do not complete a triangle or quadrangle.  We then reformulate the same search so that forbidden choices can be propagated earlier, equivalent choices can be removed by symmetry, and independent portions of the tree can be searched in parallel.

\subsection{Previous computational work}

The missing Moore graph has attracted substantial structural and computational attention.  Dalf\'o's survey~\cite{dalfo2019survey} summarizes spectral, distance-regular, and automorphism properties known for the hypothetical graph.  Ma\v{c}aj and \v{S}ir\'a\v{n}~\cite{macaj2010search} combine spectral, group-theoretic, combinatorial, and computational methods and, among other restrictions, show that the automorphism group would have order at most 375.  Such results are important to the present work because they illustrate a recurring theme: useful progress often comes from turning global existence into a more constrained representation.

Smith and Montemanni take a different computational route.  They formulate the degree-$57$ problem as an edge-maximization problem and use heuristic optimization to search for large partial graphs~\cite{smith2023optimization}.  Their later work fixes the radius-two tree and studies restricted potential subgraphs using permutation variables and Google OR-Tools CP-SAT, exploiting a cyclic permutation assumption to make the model tractable~\cite{smith2024subgraphs}.  Most recently, they showed that a construction in which the relevant derangements are drawn solely from a cyclic group is impossible~\cite{smith2026cyclic}.  These are valuable but different computational questions: heuristic proximity to a full graph, exact restricted substructures, and the exclusion of a particular algebraic family.

Our focus is on the unrestricted rooted permutation formulation itself.  We performed a targeted search of the literature, supplementary material, and public code repositories for an exhaustive implementation public timing baseline directly comparable to our solver described in the following. We did not identify one. While this prevents a non-trivial claim that our human baseline and final program is globally ``the fastest Moore-graph search program'', all performance claims in this paper are sequential internal controlled comparisons of implementations that search the same exact space, and even the presented starting point is a \emph{human-developed baseline} that in itself is \emph{non-trivial} to develop.

\subsection{Contributions}
The main contributions are the following.
\begin{enumerate}[leftmargin=*,itemsep=2pt]
\item We give a self-contained derivation of the rooted permutation representation of a diameter-two Moore graph and show how the no-triangle/no-quadrangle condition turns graph construction into a compact exact constraint problem.
\item We document a human-developed baseline implementation, including the normalizations and incremental short-cycle tests that make the naive permutation enumeration practical at small degrees.
\item We develop, in a controlled sequence, an exact CSP-style solver with reversible forbid counters, bit-vector domains, minimum-remaining-value branching, singleton propagation, and matching-based \textsc{AllDifferent} filtering.
\item We introduce a balanced, replayable parallel frontier and two increasingly strong forms of exact residual symmetry reduction: candidate-orbit pruning and canonical frontier deduplication.  These techniques reduce the $k=9$ worker tree by about 60\% relative to the balanced solver without the new symmetry layer.
\item We present a validation strategy with independent reconstruction checks and three compile-time checking levels, together with low-overhead status and depth-profile instrumentation.  The latter is useful not only operationally but also for understanding where an exact search spends its work.
\item We separate algorithmic reductions from hardware tuning and report search work, canonical frontier size, frontier-build cost, repeated wall-clock measurements, and peak resident memory.  This exposes two practical facts: fewer DFS nodes need not mean lower wall time, and even $k=57$ remains modest in memory while its subtrees become extremely expensive.
\item Finally, we report the development process itself as an algorithm-engineering case study in human-guided AI assistance: AI-generated optimization proposals are treated as hypotheses, while mathematical reasoning, exhaustive regressions, independent verification, and measured A/B tests determine which changes are retained.
\end{enumerate}

\section{Background and Human-Developed Baseline}
\label{sec:background}

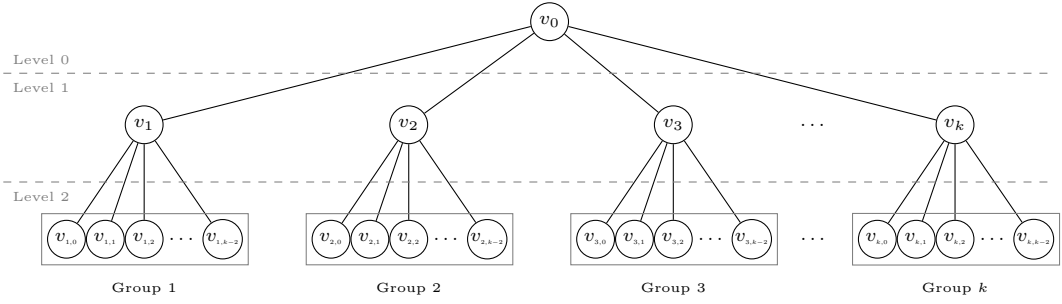
\begin{figure*}[hbt!]
\centering
\begin{tikzpicture}[scale=.78, every node/.style={font=\scriptsize}]
\node[circle,draw,inner sep=0.5pt,minimum size=5mm] (v0) at (0,0) {$v_0$};
\node[circle,draw,inner sep=0.5pt,minimum size=5mm, below left = 1cm and 5cm of v0] (v1) {$v_1$};
\node[circle,draw,inner sep=0.5pt,minimum size=5mm, below left = 1cm and 1.5cm of v0] (v2) {$v_2$};
\node[circle,draw,inner sep=0.5pt,minimum size=5mm, below left = 1cm and -2cm of v0] (v3) {$v_3$};
\node[circle,draw,inner sep=0.5pt,minimum size=5mm, below right = 1cm and 5cm of v0] (vk) {$v_k$};
\node at ($(v3)!0.5!(vk)$) {\ldots};
\path[draw] (v0) -- (v1) (v0) -- (v2) (v0) -- (v3) (v0) -- (vk);

\begin{scope}[local bounding box=group1box]
\node[circle,draw,inner sep=0.5pt,minimum size=5mm, below = 1cm of v1] (v12) {$v_{\scaleto{1,2}{2pt}}$};
\node[circle,draw,inner sep=0.5pt,minimum size=5mm, left = 0mm of v12] (v11) {$v_{\scaleto{1,1}{2pt}}$};
\node[circle,draw,inner sep=0.5pt,minimum size=5mm, left = 0mm of v11] (v10) {$v_{\scaleto{1,0}{2pt}}$};
\node[circle,inner sep=0.5pt,minimum size=5mm, right = 0mm of v12] (v13) {\ldots};
\node[circle,draw,inner sep=0.5pt,minimum size=5mm, right = 0mm of v13] (v14) {$v_{\scaleto{1,k-2}{2pt}}$};
\end{scope}
\draw (v1) -- (v10) (v1) -- (v11) (v1) -- (v12) (v1) -- (v14);
\draw[gray] ($(group1box.south west) - (1mm,1mm)$) rectangle ($(group1box.north east) + (1mm,1mm)$);

\begin{scope}[local bounding box=group2box]
\node[circle,draw,inner sep=0.5pt,minimum size=5mm, below = 1cm of v2] (v22) {$v_{\scaleto{2,2}{2pt}}$};
\node[circle,draw,inner sep=0.5pt,minimum size=5mm, left = 0mm of v22] (v21) {$v_{\scaleto{2,1}{2pt}}$};
\node[circle,draw,inner sep=0.5pt,minimum size=5mm, left = 0mm of v21] (v20) {$v_{\scaleto{2,0}{2pt}}$};
\node[circle,inner sep=0.5pt,minimum size=5mm, right = 0mm of v22] (v23) {\ldots};
\node[circle,draw,inner sep=0.5pt,minimum size=5mm, right = 0mm of v23] (v24) {$v_{\scaleto{2,k-2}{2pt}}$};
\end{scope}
\draw (v2) -- (v20) (v2) -- (v21) (v2) -- (v22) (v2) -- (v24);
\draw[gray] ($(group2box.south west) - (1mm,1mm)$) rectangle ($(group2box.north east) + (1mm,1mm)$);

\begin{scope}[local bounding box=group3box]
\node[circle,draw,inner sep=0.5pt,minimum size=5mm, below = 1cm of v3] (v32) {$v_{\scaleto{3,2}{2pt}}$};
\node[circle,draw,inner sep=0.5pt,minimum size=5mm, left = 0mm of v32] (v31) {$v_{\scaleto{3,1}{2pt}}$};
\node[circle,draw,inner sep=0.5pt,minimum size=5mm, left = 0mm of v31] (v30) {$v_{\scaleto{3,0}{2pt}}$};
\node[circle,inner sep=0.5pt,minimum size=5mm, right = 0mm of v32] (v33) {\ldots};
\node[circle,draw,inner sep=0.5pt,minimum size=5mm, right = 0mm of v33] (v34) {$v_{\scaleto{3,k-2}{2pt}}$};
\end{scope}
\draw (v3) -- (v30) (v3) -- (v31) (v3) -- (v32) (v3) -- (v34);
\draw[gray] ($(group3box.south west) - (1mm,1mm)$) rectangle ($(group3box.north east) + (1mm,1mm)$);

\begin{scope}[local bounding box=groupkbox]
\node[circle,draw,inner sep=0.5pt,minimum size=5mm, below = 1cm of vk] (vk2) {$v_{\scaleto{k,2}{2pt}}$};
\node[circle,draw,inner sep=0.5pt,minimum size=5mm, left = 0mm of vk2] (vk1) {$v_{\scaleto{k,1}{2pt}}$};
\node[circle,draw,inner sep=0.5pt,minimum size=5mm, left = 0mm of vk1] (vk0) {$v_{\scaleto{k,0}{2pt}}$};
\node[circle,inner sep=0.5pt,minimum size=5mm, right = 0mm of vk2] (vk3) {\ldots};
\node[circle,draw,inner sep=0.5pt,minimum size=5mm, right = 0mm of vk3] (vk4) {$v_{\scaleto{k,k-2}{2pt}}$};
\end{scope}
\draw (vk) -- (vk0) (vk) -- (vk1) (vk) -- (vk2) (vk) -- (vk4);
\draw[gray] ($(groupkbox.south west) - (1mm,1mm)$) rectangle ($(groupkbox.north east) + (1mm,1mm)$);

\node[below=.2cm of v12] {\tiny Group 1};
\node[below=.2cm of v22] {\tiny Group 2};
\node[below=.2cm of v32] {\tiny Group 3};
\node[below=.2cm of vk2] {\tiny Group $k$};

\node at ($(v34)!0.5!(vk0)$) {\ldots};

\node at (v0-|group1box.west) (l0) {};
\node at (v0-|groupkbox.east) (r0) {};
\node at (v1-|group1box.west) (l1) {};
\node at (vk-|groupkbox.east) (r1) {};

\draw[gray,dashed] ($(l0)!0.5!(l1)$)+(-.75cm,0)  node[above,anchor=south west] {\tiny Level 0} node[below,anchor=north west] {\tiny Level 1} -- ($(r0)!0.5!(r1)$);
\draw[gray,dashed] ($(l1)!0.5!(group1box.west)$)+(-.75cm,0) node[below,anchor=north west] {\tiny Level 2} -- ($(r1)!0.5!(groupkbox.east)$);

\end{tikzpicture}
\caption{Rooted Moore-tree representation. Remaining edges are perfect matchings between vertices in (all) pairs $ij$ of groups $1\leq i<j\leq k$.}
\label{fig:rooted-moore-tree}
\end{figure*}

We consider $k$-regular graphs of diameter two, letting $n$ denote graph order. We consistently utilize the tree-like rooted graph representation illustrated in \figref{rooted-moore-tree}, which fixes a (root) vertex $v_0$ of a degree-$k$, diameter-two Moore graph.

Since equality holds in~\eqref{eq:moorebound}, the breadth-first tree rooted at $v_0$ has exactly three levels;
\begin{itemize}[leftmargin=*]
\item Level 0: the root $v_0$,
\item Level 1: $k$ first-level vertices $v_1,\ldots,v_k$, and
\item Level 2: for every $v_i, i=1,\ldots,k$ at Level 1, there are exactly $k-1$ second-level vertices $v_{i,0},\ldots,v_{i,k-2}$.
\end{itemize}
Obviously, no two first-level vertices can coincide since that would force an edge between $v_0$ and a leaf node, producing a quadrangle. Therefore, these are all the $n=k^2+1$ vertices of the graph.

The tree already gives each leaf $v_{i,a}$ one incident edge, namely $v_i v_{i,a}$.  It still needs $k-1$ further neighbors. No two vertices in a (any one) group (see \figref{rooted-moore-tree}) can coincide, since that imposes a triangle, so all edges must connect vertices in different groups. Furthermore, no leaf node can have two or more incident vertices in any other group. If $v_{i,a}$ in group $i$ had two neighbors in group $j$, $i\neq j$, then those two vertices together with $v_j$ impose a quadrangle.  Since there are exactly $k-1$ other groups and exactly $k-1$ missing neighbors, $v_{i,a}$ must have \emph{one and only one} neighbor in every other group.

Thus, for every pair of groups $\{i,j\}$, the unknown edges form a perfect matching between two sets of size $k-1$. The matching can be represented by a permutation
\begin{equation}
             P_{ij}:\{0,\ldots,q-1\}\longrightarrow\{0,\ldots,q-1\},
\end{equation}
where $v_{i,a}$ is adjacent to $v_{j,P_{ij}(a)}$. Reversing the edge gives $P_{ji}=P_{ij}^{-1}=P_{ji}^T$.  The graph construction problem has therefore collapsed to choosing
\begin{equation}
             \binom{k}{2}
\end{equation}
permutations of size $k-1$ subject to the prohibition of triangles and quadrangles among the leaves. This permutation representation is also central in the recent potential-subgraph and derangement work of Smith and Montemanni~\cite{smith2024subgraphs,smith2026cyclic}, but here we use it without restricting the permutations to a prescribed cyclic group.

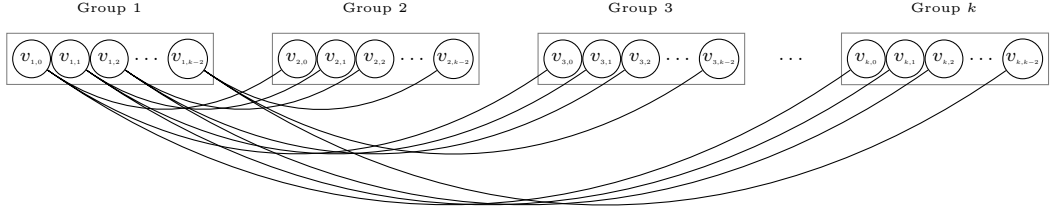
\begin{figure*}[hbt!]
\centering
\begin{tikzpicture}[scale=.78, every node/.style={font=\scriptsize}]

\begin{scope}[local bounding box=group1box]
\node[circle,draw,inner sep=0.5pt,minimum size=5mm] (v12) {$v_{\scaleto{1,2}{2pt}}$};
\node[circle,draw,inner sep=0.5pt,minimum size=5mm, left = 0mm of v12] (v11) {$v_{\scaleto{1,1}{2pt}}$};
\node[circle,draw,inner sep=0.5pt,minimum size=5mm, left = 0mm of v11] (v10) {$v_{\scaleto{1,0}{2pt}}$};
\node[circle,inner sep=0.5pt,minimum size=5mm, right = 0mm of v12] (v13) {\ldots};
\node[circle,draw,inner sep=0.5pt,minimum size=5mm, right = 0mm of v13] (v14) {$v_{\scaleto{1,k-2}{2pt}}$};
\end{scope}
\draw[gray] ($(group1box.south west) - (1mm,1mm)$) rectangle ($(group1box.north east) + (1mm,1mm)$);

\begin{scope}[local bounding box=group2box]
\node[circle,draw,inner sep=0.5pt,minimum size=5mm, right = 3cm of v12] (v22) {$v_{\scaleto{2,2}{2pt}}$};
\node[circle,draw,inner sep=0.5pt,minimum size=5mm, left = 0mm of v22] (v21) {$v_{\scaleto{2,1}{2pt}}$};
\node[circle,draw,inner sep=0.5pt,minimum size=5mm, left = 0mm of v21] (v20) {$v_{\scaleto{2,0}{2pt}}$};
\node[circle,inner sep=0.5pt,minimum size=5mm, right = 0mm of v22] (v23) {\ldots};
\node[circle,draw,inner sep=0.5pt,minimum size=5mm, right = 0mm of v23] (v24) {$v_{\scaleto{2,k-2}{2pt}}$};
\end{scope}
\draw[gray] ($(group2box.south west) - (1mm,1mm)$) rectangle ($(group2box.north east) + (1mm,1mm)$);

\begin{scope}[local bounding box=group3box]
\node[circle,draw,inner sep=0.5pt,minimum size=5mm, right = 3cm of v22] (v32) {$v_{\scaleto{3,2}{2pt}}$};
\node[circle,draw,inner sep=0.5pt,minimum size=5mm, left = 0mm of v32] (v31) {$v_{\scaleto{3,1}{2pt}}$};
\node[circle,draw,inner sep=0.5pt,minimum size=5mm, left = 0mm of v31] (v30) {$v_{\scaleto{3,0}{2pt}}$};
\node[circle,inner sep=0.5pt,minimum size=5mm, right = 0mm of v32] (v33) {\ldots};
\node[circle,draw,inner sep=0.5pt,minimum size=5mm, right = 0mm of v33] (v34) {$v_{\scaleto{3,k-2}{2pt}}$};
\end{scope}
\draw[gray] ($(group3box.south west) - (1mm,1mm)$) rectangle ($(group3box.north east) + (1mm,1mm)$);

\begin{scope}[local bounding box=groupkbox]
\node[circle,draw,inner sep=0.5pt,minimum size=5mm, right = 3.5cm of v32] (vk2) {$v_{\scaleto{k,2}{2pt}}$};
\node[circle,draw,inner sep=0.5pt,minimum size=5mm, left = 0mm of vk2] (vk1) {$v_{\scaleto{k,1}{2pt}}$};
\node[circle,draw,inner sep=0.5pt,minimum size=5mm, left = 0mm of vk1] (vk0) {$v_{\scaleto{k,0}{2pt}}$};
\node[circle,inner sep=0.5pt,minimum size=5mm, right = 0mm of vk2] (vk3) {\ldots};
\node[circle,draw,inner sep=0.5pt,minimum size=5mm, right = 0mm of vk3] (vk4) {$v_{\scaleto{k,k-2}{2pt}}$};
\end{scope}
\draw[gray] ($(groupkbox.south west) - (1mm,1mm)$) rectangle ($(groupkbox.north east) + (1mm,1mm)$);

\node[above=.2cm of v12] {\tiny Group 1};
\node[above=.2cm of v22] {\tiny Group 2};
\node[above=.2cm of v32] {\tiny Group 3};
\node[above=.2cm of vk2] {\tiny Group $k$};

\node at ($(v34)!0.5!(vk0)$) {\ldots};

\draw[out=-35,in=-145] (v10) to (v20) (v10) to (v30) (v10) to (vk0);
\draw[out=-35,in=-145] (v11) to (v21) (v11) to (v31) (v11) to (vk1);
\draw[out=-35,in=-145] (v12) to (v22) (v12) to (v32) (v12) to (vk2);
\draw[out=-35,in=-145] (v14) to (v24) (v14) to (v34) (v14) to (vk4);

\end{tikzpicture}
\caption{Symmetry breaking, first step. Group 1 can, without loss of generality, be fully defined using the identity permutation.}
\label{fig:group_1_identity_permutation}
\end{figure*}

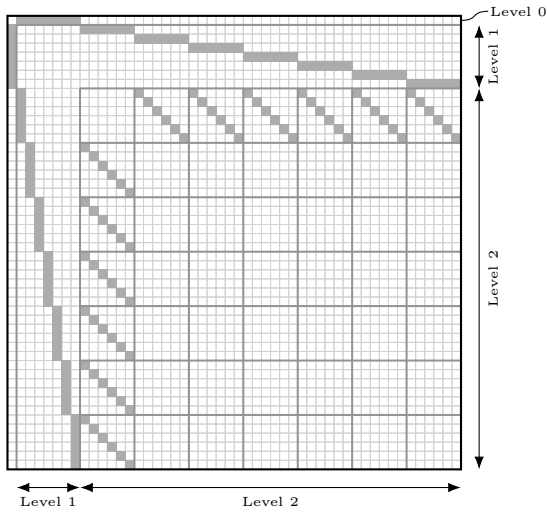
\begin{figure}[htb!]
\centering
\begin{tikzpicture}[tips=proper]

\node (M) {};

\foreach \r in {1,...,50} {
  \foreach \c in {1,...,50} {
    \wc{\r}{\c}
  }
}

\foreach \i in {2,...,8} {
  \bc{1}{\i}
  \bc{\i}{1}
}

\foreach \i in {2,...,8} {
  \foreach \j in {1,...,6} {
    \bc{\i}{8 + 6*(\i - 2) + \j}
    \bc{8 + 6*(\i - 2) + \j}{\i}
  }
}

\pidentities
\level{2}
\level{9}
\pgrid

\draw[thick] (M) rectangle ++(50*\cellsize,-50*\cellsize);

\draw[<->,>=Latex] ($(M)+(52*\cellsize,-8*\cellsize)$) edge node[rotate=90,below] {\tiny Level 2} ($(M)+(52*\cellsize,-50*\cellsize)$);
\draw[<->,>=Latex] ($(M)+(8*\cellsize,-52*\cellsize)$) edge node[below] {\tiny Level 2} ($(M)+(50*\cellsize,-52*\cellsize)$);

\draw[<->,>=Latex] ($(M)+(52*\cellsize,-1*\cellsize)$) edge node[rotate=90,below] {\tiny Level 1} ($(M)+(52*\cellsize,-8*\cellsize)$);
\draw[<->, >=Latex] ($(M)+(1*\cellsize,-52*\cellsize)$) edge node[below] {\tiny Level 1} ($(M)+(8*\cellsize,-52*\cellsize)$);

\pgfmathsetmacro{\halfcellsize}{0.5*\cellsize}
\draw ($(M)+(50*\cellsize,-\halfcellsize)$) .. controls ($(M)+(52*\cellsize,-\halfcellsize)$) and ($(M)+(51*\cellsize,\halfcellsize)$) .. ($(M)+(53*\cellsize,\halfcellsize)$) node[right,xshift=-1mm] {\tiny Level 0};

\end{tikzpicture}
\caption{Initial partial adjacency matrix structure illustrated for the $k=7$ Hoffman-Singleton graph.}
\label{fig:adjacendy_matrix_structure}
\end{figure}

We can furthermore, without loss of generality, assume that all the edges from group 1 are fully determined and represented by the identity permutations $P_{1j}=\mathit{I}_{k-1}=P_{j1}=$ for $j=1,\ldots,k$. First vertex is group 1 connects to all first vertices in the other groups, and so on. This is illustrated in \figref{group_1_identity_permutation}, and \figref{adjacendy_matrix_structure} shows the structure of the corresponding adjacency matrix.

Note that the Level 2 squares in \figref{adjacendy_matrix_structure} each correspond to a permutation matrix $P_{ij}$, representing edges between vertices in some pair of groups $i,j=1,\ldots,k$, with the diagonal $P_{ii}$ being all-zero matrices.

This is precisely the structure that is utilized and implemented in the human-developed baseline.

\subsection{Why triangles and quadrangles are enough}
The search never has to test the diameter explicitly.  The following elementary observation is the reason.

\paragraph{Proposition.}
Let $G$ be a $k$-regular graph on $n=k^2+1$ vertices.  Then $G$ is a diameter-two Moore graph if and only if it has girth five.

\paragraph{Proof.}
If $G$ is a Moore graph, equality in the breadth-first bound forbids collisions within distance two, hence forbids cycles of length three or four. Furthermore, any edge between leaf nodes imposes a cycle of length five. Conversely, if the girth is five, then from any vertex $v_0$ its $k$ neighbors are distinct and their $k-1$ remaining neighbors are all distinct from each other and from the vertices on the earlier levels.  This already accounts for $1+k+k(k-1)=k^2+1$ vertices, which is the whole graph. Every vertex is therefore within distance two of $v_0$.

This proposition explains the baseline algorithm. After fixing the degree by the perfect-matching representation, we only need to prevent short cycles. A search node is valid precisely while the edges chosen so far do not force a $C_3$ or $C_4$.

\subsection{Constraint viewpoint}

There are
\begin{equation}
  M=\binom{k}{2}(k-1)
  \label{eq:variables}
\end{equation}
row variables, one for each row of each permutation matrix.  For $k=9$, $M=288$; for $k=10$, $M=405$; and for $k=57$, $M=89{,}376$. A variable $(i,j,a)$ chooses the column $b=P_{ij}(a)$. Its raw domain is a set of $q=k-1$ values.  Each block must be a permutation, giving an \textsc{AllDifferent} constraint over its rows, while short-cycle conditions couple variables belonging to different blocks.

The target degree $57$ has an implementation-friendly accident: $q=56<64$.  Every row domain therefore fits in one 64-bit machine word.  Membership, deletion, intersection, population count, and extraction of the next candidate can all be implemented with a handful of integer instructions.  This is not mathematically deep, but it matters greatly in a search that may execute billions of such operations.

\section{Human-Developed Baseline Implementation}
\label{sec:baseline}
The original program was written before the CSP formulation in Section~\ref{sec:background} was made explicit. Nevertheless, it already exploited most of the graph theory that makes a direct search possible, and avoids the enumerate-first-then-test paradigm, constructing objects only objects that could still become Moore graphs. That is, the search space is pruned at every point at which the Moore graph structure rules are broken.

\subsection{Fixing the Moore Tree and Searching Only Matchings}
The root $v_0$, its $k$ children $v_1,\ldots,v_k$, and all $k(k-1)$ leaves are fixed once and for all. The program searches only the cross-group edges. Since every group pair is a perfect matching, a state can be stored as a triangular array of permutation rows rather than as a full adjacency matrix.

This removes several sources of waste.  Degree regularity is automatic, the order is automatic, and edges within a group are never considered.  Most importantly, the program never spends time on graphs that fail the radius-two Moore count for obvious reasons.

\subsection{Incremental Short-Cycle Rejection}
After an edge is proposed, the baseline asks whether it completes a triangle or quadrangle. The test is local. A triangle exists if $x$ and $y$ already have a common neighbor.  A quadrangle exists if a neighbor of $x$ is already adjacent to a neighbor of $y$.  If either condition holds, the candidate edge is rejected immediately. Otherwise the edge is committed and search continues.

The important point is the timing. Obviously, testing girth after a complete collection of permutations would be hopeless because almost every complete assignment would fail. Testing after each edge turns girth five from a final property into a pruning rule. The human baseline thus embodies the main mathematical and algorithmic insights of the whole project: \emph{continue only where the partial structure is Moore-capable}.

A partial conclusion here is that it is indeed possible to efficiently\footnote{The term efficiently here is taken to mean direct exclusion of all non-conforming graphs.} implement structure-aware incremental graph search in this way, at least when the structural rules are as simple as they are for Moore graphs.

\subsection{Symmetry Normalization}
A rooted Moore graph still has large relabeling freedom.  Groups can be renamed and leaf labels can be changed simultaneously without altering the underlying graph. The baseline fixes a convenient representative before searching. In permutation notation, the blocks $P_{1j}$ are fixed to the identity, and a prefix of the next block row is fixed by first-use/lexicographic conventions. These choices choose labels within an isomorphism class but do not completely remove non-isomorphic graphs.

Normalization has a clear benefit in this problem. Not only does it remove duplicate final graphs, but also large symmetric subtrees near the root of the search, where every eliminated branch would otherwise have many descendants.

\subsection{Implementation Choices}
Several elementary implementation choices were already important in the original human-developed baseline C program. State variables based on adjacency matrix entries were held in integer arrays for efficiency; reverse lookup was maintained rather than recomputed; assigned values were undone in place during backtracking; and cycle checks used the partially constructed adjacency rather than reconstructing a graph at each node.  The search order was deterministic, which made failures reproducible and simplified comparison between versions.

The baseline is therefore not a straw man. It is a specialized exact algorithm derived directly from the Moore-tree representation.
What can be claimed and measured is that this baseline provides a strong common starting point against which every subsequent change can be tested, which, at least, increases the confidence in the correctness of the final program.

\section{Human-Guided AI Improvements}
\label{sec:improvements}
The development described in this section was human-guided but AI-assisted.  The distinction matters.  An AI system can rapidly suggest familiar ideas from constraint programming, symmetry breaking, parallel search, and low-level optimization.  It cannot establish that an apparently plausible symmetry rule preserves completeness, nor can it determine whether an expensive propagator is beneficial on this particular search.  We therefore treated each proposed change as a hypothesis.  A change was retained only after (i) its correctness argument was understood, (ii) $k=7$ still produced the known Moore graph, (iii) $k=8$ still gave exhaustive nonexistence, and, when relevant, (iv) a controlled $k=9$ A/B test improved wall-clock time or materially reduced the search tree without excessive overhead.

\begin{figure*}[ht!]
\centering
\begin{tikzpicture}[node distance=6mm and 6mm,
 box/.style={draw,rounded corners,align=center,minimum height=9mm,minimum width=28mm,font=\scriptsize},
 arrow/.style={-Latex,thick}]
\node[box] (a) {\textbf{Human Baseline}\\Permutations + $C_3/C_4$ Rejection};
\node[box,right=of a] (b) {Reversible Domains\\Forbid Counters + Bitsets};
\node[box,right=of b] (c) {MRV buckets\\Singleton Propagation};
\node[box,right=of c] (d) {Matching-Based\\\textsc{AllDifferent}};
\node[box,below= of a] (e) {Replayable Balanced\\Parallel Frontier};
\node[box,right=of e] (f) {Residual Orbits\\Partition Refinement};
\node[box,right=of f] (g) {Canonical Frontier\\Deduplication};
\node[box,right=of g] (h) {Validation + Status\\Depth Profiling};
\node (dummy) at ($(d.south)!0.5!(e.north)$) {};
\draw[arrow] (a)--(b);
\draw[arrow] (b)--(c);
\draw[arrow] (c)--(d);
\draw[arrow] (d.south) -- (d.south |- dummy) -- (e.north |- dummy) -- (e.north);
\draw[arrow] (e)--(f);
\draw[arrow] (f)--(g);
\draw[arrow] (g)--(h);
\end{tikzpicture}
\caption{The retained development path.  The final solver is not the result of enabling every plausible optimization.  Several more aggressive propagation, canonicalization, and profiling variants were rejected after they slowed the program or lacked a sufficiently clear correctness argument.}
\label{fig:pipeline}
\end{figure*}
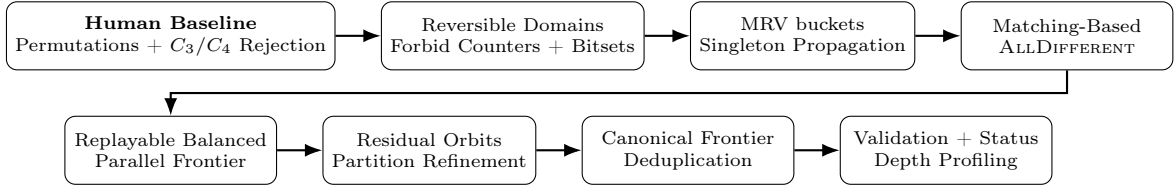

Figure~\ref{fig:pipeline} gives the development sequence.  The subsections below explain the main retained steps.

\subsection{Reversible Domains and Short-Cycle Forbid Counters}
The baseline asks whether a candidate edge is legal when the search reaches it. The first major reformulation asks the inverse question earlier: \emph{which values have already become impossible?}

For every row variable we maintain a 64-bit domain mask.  Whenever an assigned edge would make another possible edge complete a triangle or quadrangle, that value is forbidden immediately.  A single value can be forbidden for several independent reasons, so a Boolean flag is not sufficient.  We maintain a small counter
\begin{equation}
       F[x,b] = 
       \begin{minipage}{5.5cm}
       number of currently active reasons forbidding value $b$ for variable $x$.
       \end{minipage}
\end{equation}
Only the transition $0\to1$ removes a bit from the domain.  During backtracking, only the transition $1\to0$ restores it.  This detail is essential. Without reference counts, undoing one reason could incorrectly re-enable a value that remains impossible for another reason.

All updates are trailed.  A search decision records enough information to  exactly reverse the assignments, used columns, domain deletions, and short-cycle counters caused below it. No complete state is copied at an ordinary DFS node.  The resulting propagation is both stronger and cheaper than repeatedly reconstructing legal candidates from the partial graph.

\subsection{Minimum Remaining Values and Singleton Propagation}
Once domains exist, a natural branching rule is to choose the most constrained variable first. For a variable $x$, let
\begin{equation}
                   d(x)=\popcount(D_x).
\end{equation}
If $d(x)=0$, the branch is contradictory. If $d(x)=1$, the assignment is forced.  Otherwise, the search chooses a variable with minimum $d(x)$.

Scanning all $M$ variables for the minimum at every node would itself become expensive. The implementation therefore maintains doubly linked buckets indexed by domain size.  A domain deletion moves a variable between buckets in constant time, and the current smallest nonempty bucket gives an MRV variable directly. Singleton propagation repeatedly consumes the bucket $d=1$ until a fixed point is reached.

This is a good example of the interaction between algorithm and representation.  The mathematical constraints create many domain deletions; the bitset representation makes those deletions cheap; and the bucket structure converts the accumulated information into a better search order.

\subsection{Permutation Consistency as \textsc{AllDifferent}}

Every block $P_{ij}$ must be a permutation.  The cheapest form of this constraint is a used-column bit mask: once one row chooses column $b$, that value disappears from all other rows in the same block.  This is equivalent to ordinary forward checking for an \textsc{AllDifferent} constraint.

A stronger test views the remaining rows and columns as a bipartite graph.  An edge $(a,b)$ exists if row $a$ may still take column $b$.  The block can be completed only if this bipartite graph has a perfect matching.  We therefore use matching feasibility as a contradiction test and, selectively, R\'egin-style filtering~\cite{regin1994filtering} to remove edges that cannot belong to any perfect matching.

Stronger propagation is not automatically faster.  During development, variants that invoked expensive matching/SCC filtering too often were dramatically slower even though they removed more values.  The retained implementation therefore applies the stronger filtering selectively and keeps the cheaper used-column and singleton mechanisms in the hot path. \secref{lessons} returns to this negative result because it is one of the clearest algorithm-engineering lessons of this study.

\subsection{Replayable Balanced Frontiers for Parallel Search}
Depth-first search produces extremely uneven subtrees.  A naive parallelization that splits at one fixed depth can assign one thread a small branch and another a branch that runs for hours. The first parallel versions exhibited precisely this behavior.

The final approach builds an adaptive frontier.  Start with one normalized state.  Repeatedly choose a frontier leaf, replay it, apply the same MRV and propagation used by the solver, and replace that leaf by its children.  Breadth-balanced splitting continues until a requested number of jobs has been produced.  A job stores only the short sequence of decisions needed to replay it from the fixed normalized root.  Worker states are therefore independent and compact, and OpenMP dynamic scheduling can distribute jobs one at a time.

There are two useful consequences.  First, the frontier is based on the actual search heuristic rather than an arbitrary syntactic depth.  Second, the frontier becomes a natural place to perform expensive symmetry processing: a few hundred or thousand shallow states can justify canonical tests that would be too costly at every DFS node.

The number of requested jobs is a tuning parameter, not a monotone improvement knob.  Too few jobs cause load imbalance; too many spend excessive time constructing/canonicalizing the frontier and can even change the shape of the worker search unfavorably.  On a six-core/twelve-thread laptop, 1,152 requested jobs was faster for $k=9$ than either 768 or 1,536, despite 1,280 producing slightly fewer worker nodes (see \tabref{hardware-tuning}).

\subsection{Residual Symmetry: From Transpositions to Full Candidate Orbits}

The initial normalization removes a large part of the label symmetry, but not all of it.  Some group and coordinate labels remain interchangeable provided they are relabeled consistently. The first additional symmetry optimization tested whether a single label transposition preserved the partial frontier state. If two candidate values were exchanged by such an automorphism, only one representative needed to be searched.

This helped substantially, but transpositions do not generate all orbit information locally.  A partial state may be preserved by a cycle such as $(2\ 3\ 4)$ without being preserved by any of the three individual transpositions.  The stronger version therefore searches for an arbitrary residual label permutation that preserves the assigned edge set and maps one candidate to another.  Candidates connected by such automorphisms form an orbit, and only one candidate per orbit is retained.

The residual automorphism search is accelerated by invariant partition refinement.  Labels are first classified by signatures derived from their incidence with already determined parts of the state.  Labels in different cells cannot be exchanged, so exact backtracking only explores mappings within unresolved cells.  This is conceptually related to the refinement ideas used in practical graph-isomorphism software~\cite{mckay2014practical}, although our action is much smaller and specialized; it is the residual simultaneous relabeling of branch/coordinate labels, not general graph canonical labeling.

For proof safety, the automorphism search has a conservative failure policy.  If a configured search cap is reached before equivalence is proved, the candidate is \emph{kept}.  Failure to prove symmetry may cost time, but it does not remove a valid branch.

\subsection{Canonical Frontier Deduplication}
Orbit pruning removes equivalent choices while splitting one frontier state.  A stronger observation is that two frontier states reached through different splitting histories may themselves be isomorphic under the residual label action.  We therefore canonicalize the assigned edge representation of each generated frontier state and retain only one copy of each canonical key.

The implementation maps active residual labels through the admissible partition cells, encodes the assigned leaf-to-leaf edges, sorts the encoded edge list, and chooses the lexicographically least complete representation found. Deduplication is exact: a hash is used for lookup, but equality is confirmed from the canonical representation rather than from the hash alone. If the canonical search cannot be completed within its conservative cap, no deduplication is performed for that state.

This is the single most important new symmetry layer in the measured $k=9$ development.  Combined with full candidate-orbit pruning, it reduces the worker search to roughly 40\% of the no-extra-symmetry tree while leaving the deep CSP engine unchanged.

\subsection{Validation and Depth Profiling}
Exact non-existence search is useful only if the implementation is trustworthy. The final program has three verification levels that can be set compile-time. Level 0 removes expensive diagnostic checks for performance experiments. Level 1 performs cheap local consistency checks during normal operation. Level 2 additionally performs more expensive state reconstruction and independent sampled validation intended for regression tests. A found graph is independently reconstructed and checked for order, degree, and girth.

For long runs, status reporting records frontier construction time, exact scheduler state (completed/running/queued jobs), nodes and assignments, recent and average throughput, current and peak resident memory, worker balance, trail-depth quantiles, and active-job ages.  The ETA logic is deliberately conservative.  Completed frontier jobs are used only after enough duration samples exist and the active wave is not markedly heavier than the completed population; otherwise the program prints an unknown low-confidence ETA rather than extrapolating a few cheap jobs.  We also deliberately avoid interpreting trail size as percentage completion: a worker can remain in a narrow trail-size range while exploring millions of siblings.

The same instrumentation supports a sampled depth-work profile.  Every fixed number of published DFS nodes, a thread attributes a batch of work to its current depth bin.  Counters are thread-local, avoiding atomics in the hot path.  The final program writes raw CSV data, a TikZ/PGFPlots plot, and a PNG image.  For large $k$ the depth bin width is chosen automatically to keep roughly $10^3$ or fewer bins; for example, $k=57$ uses width 128 for 89,376 variables.  This profile is intended both as diagnostics and as empirical evidence about where the combinatorial search spends its effort.

\subsection{What We Rejected}
\label{sec:lessons}
A useful algorithm-engineering paper should record not only what worked.  Several plausible changes were rejected.

\paragraph{Too much global propagation.}
An aggressively modified matching/R\'egin implementation turned a $k=9$ run from seconds into minutes because expensive filtering was invoked hundreds of millions of times.  A subsequent attempt to disable the expensive filtering too broadly weakened propagation so much that the DFS tree grew by nearly an order of magnitude.  The lesson is familiar but important: propagation must be evaluated by \emph{wall time}, not by the sophistication of the consistency notion.

\paragraph{Unsafe Lex-Leader Pruning.}
A stronger lexicographic canonical rule appeared to remove almost the entire $k=8$ worker tree.  The result remained ``no graph,'' but we could not prove that the lexicographically least residual representative was always compatible with the static normalization prefix. The rule was removed. In an exhaustive non-existence calculation, unexplained spectacular speedups sometimes warrant suspicion rather than celebration.

\paragraph{Symmetry Micro-Optimizations.}
Several caches and partial automorphism consistency checks preserved exactly the same $k=9$ node count but were slightly slower.  They were discarded.  Once symmetry processing is a small fraction of total time, additional machinery can easily cost more than it saves.

\paragraph{More Jobs are not Always Better.}
Increasing the frontier can reduce worker nodes while increasing wall time.  Canonicalization and replay have a cost, and the resulting jobs need not have equal difficulty.  The optimum depends on hardware, degree, and check level, so the program exposes the job count rather than hard-coding a very deep frontier.

\subsection{Measured Development}
We separate algorithmic effects from hardware effects.  The first experiment keeps the hardware configuration and frontier target fixed and changes only the symmetry layer.  Table~\ref{tab:symmetry-ablation} uses six threads and 768 requested $k=9$ frontier jobs throughout.  The final row adds full residual candidate orbits, partition refinement, and exact canonical deduplication of generated frontier states.

\begin{table}[htb!]
\centering
\caption{Controlled $k=9$ symmetry ablation with six threads and 768 requested frontier jobs.  These development runs isolate the algorithmic effect of successively stronger frontier symmetry.}
\label{tab:symmetry-ablation}
\small
\resizebox{.49\textwidth}{!}{
\begin{tabular}{lrrr}
\toprule
Frontier symmetry & Worker nodes & Assignments & Wall (sec)\\
\midrule
none & 7,293,331 & 28,537,446 & 47.6\\
transpositions & 5,820,280 & 22,689,245 & 37.2\\
full orbits\ldots  & 2,920,685 & 11,252,982 & 17.9\\
\bottomrule
\end{tabular}
}
\end{table}

Relative to the first row, the final symmetry stack removes 60.0\% of the worker nodes and 60.6\% of the assignments, while reducing wall time by about 62\%.  The close agreement is useful: canonicalization does non-trivial work, but at this shallow depth its cost is small compared with the subtrees it removes.

The second experiment asks a different question: how much frontier decomposition is useful once many hardware threads are available?  Table~\ref{tab:hardware-tuning} reports the current final solver on a 16-core machine, using both 16 and 32 hardware threads.  The 32-thread entries are medians of five complete runs; the 16-thread entries are single tuning runs.  A larger frontier is not monotonically better.  For 32 threads, 1,024 requested jobs gives the lowest measured median wall time, even though 1,280 jobs explores slightly fewer worker nodes.

\begin{table}[htb!]
\centering
\caption{$k=9$ frontier and thread tuning on the 16-core machine.  The 32-thread wall times are medians of five runs; 16-thread timings are single tuning runs.}
\label{tab:hardware-tuning}
\small
\resizebox{.49\textwidth}{!}{
\begin{tabular}{rrrrr}
\toprule
Threads & Requested & Canonical & Nodes & Wall (s)\\
\midrule
16 & 896  & 630 & 2,883,242 & 8.21\\
16 & 1,024 & 712 & 2,764,882 & 7.98\\
16 & 1,152 & 848 & 2,767,847 & \textbf{7.91}\\
16 & 1,280 & 966 & 2,759,419 & 7.98\\
\midrule
32 & 768  & 555 & 2,920,685 & 5.78\\
32 & 896  & 630 & 2,883,242 & 5.66\\
32 & 1,024 & 712 & 2,764,882 & \textbf{5.28}\\
32 & 1,152 & 848 & 2,767,847 & 5.34\\
32 & 1,280 & 966 & 2,759,419 & 5.33\\
\bottomrule
\end{tabular}
}
\end{table}

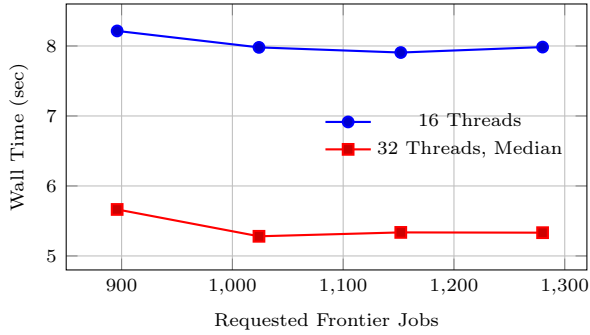
\begin{figure}[htb!]
\centering
\begin{tikzpicture}
\begin{axis}[
 width=\columnwidth,height=5.1cm,
 xlabel={Requested Frontier Jobs},
 ylabel={Wall Time (sec)},
 xmin=850,xmax=1320,ymin=4.8,ymax=8.6,
 grid=major,
 legend style={font=\scriptsize,draw=none,fill=none,at={(0.98,0.63)},anchor=north east},
 tick label style={font=\scriptsize},label style={font=\scriptsize}]
\addplot+[mark=*,thick] coordinates {(896,8.2149) (1024,7.9799) (1152,7.9063) (1280,7.9843)};
\addlegendentry{16 Threads}
\addplot+[mark=square*,thick] coordinates {(896,5.6643) (1024,5.2808) (1152,5.3364) (1280,5.3325)};
\addlegendentry{32 Threads, Median}
\end{axis}
\end{tikzpicture}
\caption{Parallel tuning is a coupled algorithm--hardware problem.  More frontier jobs improve granularity only up to the point where construction, canonicalization, replay, and a less favorable partition outweigh the scheduling benefit.}
\label{fig:hardware-tuning}
\end{figure}

On the six-core/twelve-thread laptop, six repeated $k=9$ runs with 1,152 requested jobs and validation disabled have identical search counts (2,767,847 nodes and 10,700,474 assignments) and wall times between 11.95 and 13.03 seconds, with median 12.42 seconds.  The deterministic node count combined with variable OpenMP job ownership is a useful reproducibility check: scheduling changes who performs the work, but not which search tree is explored.

\paragraph{Stress regime at $k=10$.}
The qualitative transition occurs immediately after $k=9$.  On the 16-core/32-thread machine, 768 requested jobs produce 487 canonical jobs in 0.71 seconds.  Twelve jobs finish quickly, after which all 32 workers remain inside the next wave: the completed count is still 12 after 3.5 minutes while 88.8 million worker nodes have been processed.  Frontiers of 896 and 1,024 jobs show the same pattern.  This is why the status estimator treats the active wave as censored evidence and suppresses a point ETA when completed jobs are clearly unrepresentative.

\paragraph{Stress regime at $k=57$.}
At degree 57, state size remains manageable but subtree cost becomes enormous.  With 32 threads and 512 requested jobs, frontier construction takes 24.6 seconds and yields 511 jobs; no job finishes during a 2.5-minute diagnostic run, although aggregate throughput reaches roughly $2.7\times10^5$ worker nodes/s.  A 768-job frontier is substantially less favorable in this short test, while 1024 jobs returns to a similar node rate but doubles frontier-build cost.  These are tuning observations only: no complete $k=57$ search is claimed and no completion time is extrapolated.

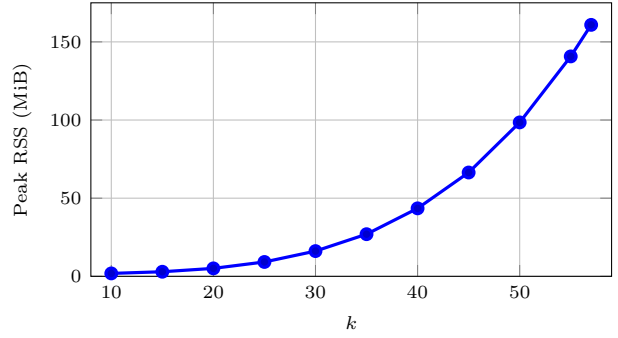
\begin{figure}[t]
\centering
\begin{tikzpicture}
\begin{axis}[
 width=\columnwidth,height=5.2cm,
 xlabel={$k$},ylabel={Peak RSS (MiB)},
 xmin=8,xmax=59,ymin=0,ymax=175,
 grid=major,
 tick label style={font=\scriptsize},label style={font=\scriptsize}]
\addplot+[mark=*,very thick] coordinates {
 (10,1.867) (15,2.930) (20,5.086) (25,9.188) (30,16.145)
 (35,26.961) (40,43.488) (45,66.422) (50,98.484) (55,140.707) (57,160.891)};
\end{axis}
\end{tikzpicture}
\caption{Peak resident memory in five-minute stress runs with 12 threads and 1152 requested jobs.  Memory growth is modest compared with the combinatorial growth of running time; even the $k=57$ process remains below 161 MiB RSS.}
\label{fig:memory-scaling}
\end{figure}

The space behavior is much more benign.  Figure~\ref{fig:memory-scaling} reports five-minute runs on the twelve-thread laptop, all with 1,152 requested jobs.  Peak resident memory rises from 1.9 MiB at $k=10$ to 160.9 MiB at $k=57$.  Thus the missing instance is not difficult because a solver state is too large to store; it is difficult because the number and cost of states that may need to be explored grow dramatically.

The profiler records a complementary quantity: where worker computation is spent as a function of trail size. It samples depth only at an already existing publication point (once per $2^{16}$ worker nodes by default), accumulates thread-local bins, and distributes each completed frontier job's measured time over its sampled bins in proportion to node work. The resulting raw profile data is output in a CSV-file, and the solver also produces a TikZ plot and a dependency-free PNG at the end of a completed run for a depth profile visualization.  Because trail size repeatedly rises and falls during DFS, the resulting profile is a workload distribution, not a percentage-complete indicator.

\section{Discussion and Future Work}
\subsection{What Became Faster, and Why?}
The largest measured gain did not come from making one inner loop a few percent cheaper.  It came from searching \emph{less}. The original rooted representation and immediate pruning already removes the enormous space of non-conforming graphs. Domain propagation then rejects impossible permutation values before they are explicitly tried.  MRV moves contradictions closer to the root.  The balanced frontier makes parallelism effective.  Finally, residual symmetry prevents large equivalent subtrees from being explored at all.

On the $k=9$ benchmark, the symmetry sequence is especially instructive.  Twelve shallow transposition equivalences already removed about 20\% of the worker tree in an early version.  Moving from transpositions to full residual orbits and canonical frontier deduplication ultimately removed roughly 60\% relative to the no-extra-symmetry balanced solver.  This is precisely where symmetry breaking should be applied: close enough to the root that one proof of equivalence saves a large subtree, but sparse enough that canonicalization does not burden every DFS node.

Hardware effects must be kept separate.  The same final $k=9$ search takes a median 12.42 seconds in six repeated runs on the six-core/twelve-thread laptop, while the 16-core machine reaches a median 5.28 seconds over five runs with 32 hardware threads and 1,024 requested jobs.  This is a hardware and parameter-tuning gain, not an algorithmic reduction: the frontier target changes the worker tree slightly.  We therefore report search work, frontier size, and wall time separately rather than folding everything into one headline speedup.

\subsection{The Role of AI Assistance}
AI assistance was valuable mainly because it shortened the proposal--implementation--measurement loop.  It quickly surfaced standard but relevant techniques---MRV, reversible domains, matching filters, symmetry orbits, canonicalization, dynamic job scheduling, and profiling ideas---and generated implementation variants that would otherwise have taken considerably longer to prototype.

The failures are equally informative.  Some AI-proposed changes were incorrect until repaired; others were correct but dramatically slower; still others produced pruning that looked attractive but lacked a sufficiently convincing completeness argument.  The working methodology was not to simply ``ask AI to optimize the program.''  It was closer to experimental mathematics and algorithm engineering: formulate a concrete hypothesis, make the invariants explicit, compare against known cases, inspect the complete search counts, and retain the change only when both correctness and performance survive scrutiny.

This use also creates an ethical and reproducibility obligation.  AI-generated code can contain subtle errors, particularly in reversible state updates and symmetry logic. The source release includes regression commands for $k=7$ and $k=8$, compile-time validation levels, benchmark scripts, and raw logs used for the tables.

\subsection{Where is the Source Code?}
You can find the original human-developed baseline C source code and the final improved code on GitHub
\iftoggle{anonymous} 
{%
\cite{anonGithub}.
}
{%
\cite{lessismooregithub}.
} 

\subsection{What Can Still be Improved?}
There is ample room for improvement, but the development experience suggests where to be cautious.

The first direction is \emph{stronger but provably compatible symmetry}.  Full canonical augmentation deeper in the DFS could in principle remove many more equivalent states, but it should be introduced only with a clean proof that the canonical representative respects the existing static normalization.  A general canonical-labeling package could also be used at sparse checkpoints, although converting the specialized permutation state to a general graph may cost more than a custom stabilizer calculation.

The second direction is \emph{learning from conflicts}.  The current solver backtracks chronologically and mostly forgets why a branch failed.  Lightweight nogoods or conflict-directed backjumping could avoid rediscovering the same local impossibility under different decision orders.  Because the state has considerable symmetry, any learned object would ideally be canonicalized or expressed in an invariant local form.

The third direction is \emph{using more graph theory before search}.  Structural results about automorphisms, spectra, induced distance-regular subgraphs, and forbidden algebraic permutation families~\cite{dalfo2019survey,macaj2010search,smith2026cyclic} may be convertible into additional exact constraints.  This is arguably the most promising route toward degree 57: not merely search the existing formulation faster, but shrink the formulation by proving that whole classes of partial structures cannot occur.

A fourth direction is methodological.  The depth profile can be extended to record MRV domain size, contradiction rate, and effective branching factor by depth.  Such measurements may reveal phase transitions in the search and help design better branching heuristics.  They may also yield better online estimates for long jobs, although we have deliberately avoided presenting an unvalidated completion-time model here.

\subsection{What Questions Might Efficient Software Plausibly Answer?}
A complete unrestricted $k=57$ search is not presently a realistic target on the hardware used here.  Efficient exact software can nevertheless contribute to Moore-graph research in more focused ways.  It can exhaustively test additional symmetry hypotheses; classify permutation systems under restricted automorphism actions; search for or exclude prescribed local configurations; enumerate feasible rooted substructures beyond current CP-SAT constructions; and test conjectured consequences of spectral or distance-regular structure.  These are finite questions for which a negative computation can be meaningful if the search space, symmetry reduction, and verifier are explicit.

This perspective changes the role of the missing Moore graph.  Instead of treating degree 57 as one enormous yes/no instance, we can use it as a source of increasingly constrained exact subproblems.  The software developed here is intended as a platform for those experiments.

The same methodology presented here is also directly useful for instance and structure discovery in problem relaxations relating to Moore graphs, for example mixed Moore graphs by Dalf\'o et al. in~\cite{dalfo2022moorelargestmixedgraphs}.

\section{Summary}
A diameter-two Moore graph is one of those objects for which a simple counting argument creates an extraordinarily rigid equality case. Fixing one vertex exposes the entire graph as a radius-two tree plus perfect matchings between its leaf groups. That observation turns the construction problem from arbitrary graph search into a system of interacting permutations, and girth five reduces validity to the avoidance of triangles and quadrangles.

Starting from that human-developed representation, we have shown how a sequence of exact implementation changes can make the search substantially smaller and faster. Reversible domains and short-cycle forbid counters propagate local graph constraints; MRV and singleton rules use the propagated information; matching enforces permutation feasibility; a replayable frontier makes parallel execution balanced; and residual orbit pruning plus canonical frontier deduplication removes equivalent shallow subtrees. On $k=9$, the strongest symmetry stack reduces the measured worker tree from 7.29 million to 2.92 million nodes on the same six-thread/768-job setup. The current implementation completes the tuned benchmark in median 12.42 seconds on the six-core/twelve-thread laptop and 5.28 seconds on the 16-core/32-thread machine. At the other end of the scale, a five-minute $k=57$ stress run peaks at only 160.9 MiB RSS, confirming that search complexity rather than memory capacity is the dominant barrier.

The results do not bring an exhaustive degree-$57$ search within reach.  The incomplete $k=10$ and $k=57$ stress tests instead make the opposite point: combinatorial growth rapidly overwhelms ordinary hardware, even after large constant-factor and search-space improvements. The value of this work is its reproducible account of how mathematical structure, exact constraint propagation, symmetry, and parallel algorithm engineering can be combined to push specialized graph search considerably, and of where the remaining difficulty begins.

\section*{Declarations}
\paragraph{AI assistance.}
Generative AI tools were used substantively during algorithm exploration, code prototyping, literature discovery, and initial drafting. All mathematical claims, citations, code changes, and experimental conclusions have been checked by the (human) authors against primary sources, exact regression tests, independent graph verification, and raw execution logs. The authors assume responsibility for all content.


\bibliographystyle{siamplain}
\bibliography{less_is_moore}

\end{document}